\documentclass[english,aps,prl,notitlepage,reprint]{revtex4-1}
\usepackage[latin9]{inputenc}
\usepackage{graphicx}
\usepackage{epsfig}

\makeatletter

\@ifundefined{textcolor}{}
{%
 \definecolor{BLACK}{gray}{0}
 \definecolor{WHITE}{gray}{1}
 \definecolor{RED}{rgb}{1,0,0}
 \definecolor{GREEN}{rgb}{0,1,0}
 \definecolor{BLUE}{rgb}{0,0,1}
 \definecolor{CYAN}{cmyk}{1,0,0,0}
 \definecolor{MAGENTA}{cmyk}{0,1,0,0}
 \definecolor{YELLOW}{cmyk}{0,0,1,0}
}

\makeatother

\begin{document}

\title{Hadronic Matter in the Robertson-Walker Metric and the Early Universe}

\author{I. E. Cunha}
\affiliation{Centro Brasileiro de Pesquisas F{\'{\i}}sicas}
\author{C. C. Barros Jr.}
\affiliation{Departamento de F{\'{\i}}sica, CFM, Universidade Federal de Santa Catarina}
\begin{abstract}
In this work, the Friedman equations for hadronic matter in the Robertson-Walker
metric in the early Universe are obtained. We consider the hadronic phase, formed 
after the hadronization of the quark-gluon plasma, that means times 
from $10^{-6}s$ to 1$s$. The set of equations will be derived and the behavior of the
system will be studied considering one
approximate analytical solution.
\end{abstract}
\maketitle

In the past few years, at CERN \cite{cern},\cite{na1} and  at RHIC \cite{rh1}-\cite{rh3} very important results
have been obtained. It has been observed that when matter is submitted to the extreme 
conditions of the high-energy heavy-ion collisions
(high temperatures, $T>$ 150 MeV, high energy densities, exceding 1 GeV/fm,  and small barionic chemical potential $\mu\sim 0$), 
it is believed that the quark-gluon plasma phase is produced. At the theoretical level
this understanding is improving \cite{gy}-\cite{bla}, and gradually, the behavior of the QCD phase diagram is being
clarified. 

These conditions are believed to be close to the ones found in the early Universe, so,
we may be tempted to try to understand the early Universe in a similar way \cite{abd}-\cite{raf2}. In high-energy collisions, 
after the QGP creation, it expands and cools, and the hadronization occurs during this expansion, determining the 
freeze-out process \cite{bra},\cite{flo}.

The hadronic phase produced in high energy heavy-ion collisions has been successfully studied in
\cite{deb1}-\cite{bm}, in terms of effective relativistic nuclear models considering a mean-field theory
by means of a Walecka model \cite{wal}, and its main features could be explained with this procedure.
 This fact support the idea of using this model, at least as a first approximation, in this work.

On the other hand, one must consider hot hadronic matter in a curved space-time.
 In \cite{ruf1},\cite{ruf2} this kind of system has been considered, but for studying
the neutron stars structure ($T\sim$0).

So, in this work, we investigate the evolution of the early Universe, by the Friedman 
equations, considering an effective relativistic mean-field model for the hadronic phase, with
the strong interactions given by the exchange of $\sigma$, $\omega$ and $\rho$ mesons. 
We suppose that this model is valid after the hadronization and before
the nuclei formation, that means times from $10^{-6}s$ up to 1$s$, approximately.
The equations will be obtained in this framework and one approximate analytical 
solution will be shown.

 To study  the evolution of the Universe we will consider the Robertson-Walker metric
\begin{equation}
ds^{2}=dt^{2}-R^{2}\left(t\right)\left[\frac{dr^{2}}{1-kr^{2}}+r^{2}d\theta^{2}+r^{2}\sin^{2}
\theta d\phi^{2}\right],
\label{rw}
\end{equation}

\noindent
that assumes an isotropic and homogeneous hadronic phase.

In this phase, we will consider the $\sigma-\omega-\rho$
model adapted to the curved space-time in a way similar to the one shown in \cite{ruf2}, but now,
considering the Robertson-Walker metric (\ref{rw}). The Lagrangian  density is
\begin{equation}
\mathcal{L}=\mathcal{L}_{b}+\mathcal{L}_{e}+\mathcal{L}_{\nu}+\mathcal{L}_{\sigma}+\mathcal{L}_{\omega}+\mathcal{L}_{\rho}+\mathcal{L}_{\gamma}+\mathcal{L}_{int}+\mathcal{L}_{g}  \  ,
\end{equation}

\noindent
where
the first term, $\mathcal{L}_{b}$,
describes the baryons (protons and neutrons) and, in the same way,
the terms $\mathcal{L}_{e}$ and $\mathcal{L}_{\nu}$ are related to the
electrons and neutrinos respectively. The term $\mathcal{L}_{\sigma}$
refers to the Lagrangian densitiy of the free-scalar boson field $\sigma$, which describes 
the long range attractive interaction. The terms $\mathcal{L}_{\omega}$
and $\mathcal{L}_{\rho}$ are the Lagrangians of the free massive vector boson fields $\omega$ and $\rho$, 
that describe the short range interactions. Now, the term
$\mathcal{L}_{\gamma}$ is the Lagrangian of the electromagnetic field.
Finally, $\mathcal{L}_{int}$ represents the interacting part of the Lagrangian density and the term $\mathcal{L}_{g}$ 
is the Lagrangian of the free gravitational field. The quantities $g_{\sigma}$, $g_2$, $g_3$,
$g_{\omega}$ and $g_{\rho}$ are the coupling constants, and can be found in the in the references \cite{nl3}-\cite{tm3}, 
$\sigma_3$ is the Pauli matrix, and $e$ is the proton charge. Other observation is that $\nabla_\mu$ 
is the covariant derivative of the respective object that it acts.
These Lagrangian densities may be expressed as
\begin{eqnarray}
\mathcal{L}_{b} & = & \overline{\Psi}\left(i\gamma^{\mu}\nabla_{\mu}-\widetilde{M}_{b}\right)\Psi,\\
\mathcal{L}_{e} & = & \overline{\psi_{e}}\left(i\gamma^{\mu}\nabla_{\mu}-m_{e}\right)\psi_{e},\\
\mathcal{L}_{\sigma} & = & \frac{1}{2}\nabla_{\mu}\sigma\nabla^{\mu}\sigma-\frac{1}{2}m_{\sigma}^{2}\sigma^{2}-U\left(\sigma\right),\\
\mathcal{L}_{\omega} & = & -\frac{1}{4}\Omega_{\mu\nu}\Omega^{\mu\nu}+\frac{1}{2}m_{\omega}^{2}\omega_{\mu}\omega^{\mu},\\
\mathcal{L}_{\rho} & = & -\frac{1}{4}\Xi_{\mu\nu}\Xi^{\mu\nu}+\frac{1}{2}m_{\rho}^{2}\rho_{\mu}\rho^{\mu},\\
\mathcal{L}_{\gamma} & = & -\frac{1}{4}F_{\mu\nu}F^{\mu\nu},\\
\mathcal{L}_{int} & = & -g_{\omega}\overline{\Psi}\gamma^{\mu}\omega_{\mu}\Psi-\frac{1}{2}g_{\rho}\overline{\Psi}\gamma^{\mu}\sigma_{3}\rho_{\mu}\Psi\nonumber\\
& & -e\overline{\Psi}\gamma^{\mu}\left(\frac{1+\sigma_{3}}{2}\right)A_{\mu}\Psi +e\overline{\psi_{e}}\gamma^{\mu}A_{\mu}\psi_{e},\\
\mathcal{L}_{g} & = & -\frac{R}{16\pi G}
\end{eqnarray}

\noindent
with
\begin{eqnarray}
\widetilde{M}_{b} & = & \left(\begin{array}{cc}
m_{p}-g_{\sigma}\sigma & 0\\
0 & m_{n}-g_{\sigma}\sigma
\end{array}\right)\\
&=&\left(\begin{array}{cc}
\widetilde{m}_{p} & 0\\
0 & \widetilde{m}_{n}
\end{array}\right),\\
\Omega_{\mu\nu} & = & \partial_{\mu}\omega_{\nu}-\partial_{\nu}\omega_{\mu},\\
\Xi_{\mu\nu} & = & \partial_{\mu}\rho_{\nu}-\partial_{\nu}\rho_{\mu},\\
F_{\mu\nu} & = & \partial_{\mu}A_{\nu}-\partial_{\nu}A_{\mu},\\
\Psi & = & \left(\begin{array}{c}
\psi_{p}\\
\psi_{n}
\end{array}\right),\quad p\rightarrow proton,\; n\rightarrow neutron\\
U\left(\sigma\right) & = & \frac{1}{3}g_{2}\sigma^{3}+\frac{1}{4}g_{3}\sigma^{4}.
\end{eqnarray}

Two observations must be made about this formulation: we will consider the difference of mass 
between the proton and the neutron and we will not show the neutrino's Lagrangian. The reason to
 consider the difference of mass is because it causes a discrepancy between the neutron
and proton densities that we desire to capture with the model. Now, the reason for not showing the Lagrangian 
of the neutrinos is just because we will not be concerned with a model for neutrinos in a first approach, 
this study deals with hadronic interactions in the hadronic phase and
we can approximate its contribution as an ultra-relativistic Fermi gas as it will be made below. 
The spin connections that appears in the fermion's covariant derivatives, that should be considered in this kind 
of picture, will be neglected in this formulation as its effect is negligible in a mean field theory, 
and this study will be left as a refinement for future works.

Now if we define
\begin{eqnarray}
\left\langle \overline{\Psi}\Psi\right\rangle  & = & \rho_{s},\\
\left\langle \overline{\Psi}\gamma^{\mu}\Psi\right\rangle  & = & J_{b}^{\mu}=J_{p}^{\mu}+J_{n}^{\mu},\\
\left\langle \overline{\Psi}\gamma^{\mu}\tau_{3}\Psi\right\rangle  & = & J_{3}^{\mu}=J_{p}^{\mu}-J_{n}^{\mu},\\
\left\langle \overline{\Psi}\gamma^{\mu}\left(\frac{1+\tau_{3}}{2}\right)\Psi\right\rangle  & = & J_{p}^{\mu},\\
\left\langle \overline{\psi_{e}}\gamma^{\mu}\psi_{e}\right\rangle  & = & J_{e}^{\mu},\\
\left\langle \sigma\right\rangle  & = & \phi,\\
\left\langle \Omega^{\mu\nu}\right\rangle =V^{\mu\nu}&,&\left\langle \omega^{\mu}\right\rangle =V^{\mu},\\\left\langle \Xi^{\mu\nu}\right\rangle  = B^{\mu\nu}&,&\left\langle \rho^{\mu}\right\rangle =B^{\mu},\\\left\langle F^{\mu\nu}\right\rangle  = F^{\mu\nu}&,&\left\langle A_{\mu}\right\rangle = A_{\mu},
\end{eqnarray}

\noindent
we may
derive the equations of motion by applying the mean-field approximation
\begin{eqnarray}
\frac{1}{\sqrt{-g}}\partial_{\mu}\left(\sqrt{-g}\partial^{\mu}\phi\right) & = & g_{\phi}\rho_{s}-m_{\phi}^{2}\phi - \frac{\partial U}{\partial\phi},\label{eq:aa}\\
\frac{1}{\sqrt{-g}}\partial_{\mu}\left(\sqrt{-g}V^{\mu\nu}\right) & = & g_{V}J_{b}^{\nu}-m_{V}^{2}V^{\nu} \nonumber \\
&=& g_{V}n_{b}u^{\nu}-m_{V}^{2}V^{\nu} ,\\
\frac{1}{\sqrt{-g}}\partial_{\mu}\left(\sqrt{-g}B^{\mu\nu}\right) & = & \frac{1}{2}g_{B}\left(J_{p}^{\nu}-J_{n}^{\nu}\right)-m_{B}^{2}B^{\nu}\nonumber \\
& = & \frac{1}{2}g_{B}\left(n_{p}-n_{n}\right)u^{\nu} -m_{B}^{2}B^{\nu}, \;\; \\
\frac{1}{\sqrt{-g}}\partial_{\mu}\left(\sqrt{-g}F^{\mu\nu}\right) & = & e\left(J_{p}^{\nu}-J_{e}^{\nu}\right)\nonumber \\
& = & e\left(n_{p}-n_{e}\right)u^{\nu}, \label{eq:bb}
\end{eqnarray}

\noindent
where $u^{\mu}$ is the average quadri-velocity of the charge carriers,
and the densities are
\begin{eqnarray}
\rho_{s} & = & \sum_{p,n}\frac{2}{\left(2\pi\right)^{3}}\int d^{3}k\frac{\widetilde{m}_{i}}{\sqrt{k^{2}+\widetilde{m}_{i}^{2}}}\left(f_{i+}+f_{i-}\right),\\
n_{b} & = & \sum_{p,n}\frac{2}{\left(2\pi\right)^{3}}\int d^{3}k\left(f_{i+}-f_{i-}\right)=n_{p}+n_{n},\\
n_{e} & = & \frac{2}{\left(2\pi\right)^{3}}\int d^{3}k\left(f_{e+}-f_{e-}\right),
\end{eqnarray}

\noindent
where
\begin{eqnarray}
f_{i\pm} & = & \frac{1}{e^{\beta\left(E_{i}\mp\nu_{i}\right)}+1},\quad i=p,n\\
f_{e\pm} & = & \frac{1}{e^{\beta\left(E_{e}\mp\nu_{e}\right)}+1},
\end{eqnarray}

\noindent
that are the Fermi-Dirac distributions with index
``$+$'' for particles and ``$-$'' for anti-particles, $\nu_{i}$
and $\nu_{e}$ are the effective chemical potentials that have the form
\begin{eqnarray}
\nu_{p} & = & \mu_{p}-g_{V}V^{0}-\frac{1}{2}g_{B}B^{0},\\
\nu_{n} & = & \mu_{n}-g_{V}V^{0}+\frac{1}{2}g_{B}B^{0},\\
\nu_{e} & = & \mu_{e}
\end{eqnarray}
in a homogeneous and locally neutral system (for electric charge)
and with $u^{\mu}=\left(1,\mathbf{0}\right)$ (that is the case that we have here by virtue of the constraint imposed by the metric).
Note that the factor ``$2$'' in each density is the spin degeneracy factor
and we also have changed the index in the equations (\ref{eq:aa})-(\ref{eq:bb}) for correspondence with the field which it refers.
The equations derived above show how we can study a medium, composed by interacting hadrons, considering the mean-field approximation. In this framework, the metric is also considered, and the composition of this phase also has effect on the determination of the metric.

Now, we can apply these ideas in order to describe the metric in the hadronic phase of the Universe, that means times above $10^{-6}s$ and below 1$s$. For this purpose, as a first approximation, we will consider a isotropic and homogeneous Universe, described by the Robertson-Walker
metric (\ref{rw}), that determines
the local neutrality for electric charge. The constraints 
lead us from the equations (\ref{eq:aa})-(\ref{eq:bb}) to the set of equations
\begin{eqnarray}
\frac{d^{2}\phi}{dt^{2}}+3\frac{\dot{R}}{R}\frac{d\phi}{dt}+m_{\phi}^{2}\phi+\frac{\partial U}{\partial\phi}&=&g_{\phi}\rho_{s},\label{eq:phi}\\
m_{V}^{2}V^{0} & = & g_{V}J_{b}^{0}=g_{V}n_{b},\label{eq:V=00003Dn}\\
m_{B}^{2}B^{0} & = & g_{B}J_{3}^{0}\nonumber \\
&=&g_{B}\left(n_{p}-n_{n}\right),\label{eq:B=00003Dn}  \  \ 
\end{eqnarray}
where we suppressed the electromagnetic field equation because it vanishes if homogeneity and isotropy are imposed. Another important condition is the conservation of the baryon number that is expressed by $\nabla_{\mu}{J_b}^{\mu}=0$ and gives the result
\begin{equation}
n_{b}=n_{b0}\left(\frac{R_{0}}{R}\right)^{3}   \  ,
\end{equation}
\\
where $n_{b0}$ is the initial baryon density (when the hadron epoch begins).

Supposing that the matter may be considered as a perfect fluid, we use an energy-momentum tensor of the form $T^{\mu\nu}=\left(\epsilon+p\right)u^{\mu}u^{\nu}-pg^{\mu\nu}$ with $u^{\mu}=\left(1,\mathbf{0}\right)$. 
Making use of the Lagrangian densities determined above, we may obtain
\begin{eqnarray}
\epsilon & = & \epsilon_{\phi}+\epsilon_{V}+\epsilon_{B}+\epsilon_{e}+\epsilon_{\nu}+\epsilon_{\gamma}+\epsilon_{b}\nonumber \\
 & = & \frac{1}{2}\left(\frac{d\phi}{dt}\right)^{2}+\frac{1}{2}m_{\phi}^{2}\phi^{2}+U\left(\phi\right)\nonumber \\
 & &+\frac{1}{2}\frac{g_{V}^{2}}{m_{V}^{2}}n_{b0}^{2}\left(\frac{R_{0}}{R}\right)^{6}+\frac{1}{2}\frac{g_{B}^{2}}{m_{B}^{2}}\left(n_{p}-n_{n}\right)^{2} +\frac{43}{120}\pi^{2}T^{4}\nonumber \\
 &  & +\frac{2}{\left(2\pi\right)^{3}}\sum_{p,n}\int d^{3}k\sqrt{k^{2}+\widetilde{m}_{i}^{2}}\left(f_{i+}+f_{i-}\right),\\
p & = & p_{\phi}+p_{V}+p_{B}+p_{e}+p_{\nu}+p_{\gamma}+p_{b}\nonumber \\
 & = & \frac{1}{2}\left(\frac{d\phi}{dt}\right)^{2}-\frac{1}{2}m_{\phi}^{2}\phi^{2}-U\left(\phi\right)\nonumber \\
 & & +\frac{1}{2}\frac{g_{V}^{2}}{m_{V}^{2}}n_{b0}^{2}\left(\frac{R_{0}}{R}\right)^{6}+\frac{1}{2}\frac{g_{B}^{2}}{m_{B}^{2}}\left(n_{p}-n_{n}\right)^{2}+\frac{43}{360}\pi^{2}T^{4}\nonumber \\
 &  & +\frac{1}{3}\frac{2}{\left(2\pi\right)^{3}}\sum_{p,n}\int d^{3}k\frac{k^{2}}{\sqrt{k^{2}+\widetilde{m}_{i}^{2}}}\left(f_{i+}+f_{i-}\right).
\end{eqnarray}
\\
But is necessary to make some observations. As we said before, the electromagnetic field vanish if we use rigorously the Robertson-Walker metric. However the contribution of the photons is significant in this phase of the Universe. Then we included the photons as a boson gas,
that is a good approximation, as electromagnetic interactions plays a role of small corrections in the hadronic phase (fact that is widely used in the study of high-energy heavy-ion collisions). For similar reasons, the electrons and neutrinos are approximated by a Fermi gas without chemical potential. These approximation lead us to consider $\mu_p=\mu_n\equiv \mu_b$ in the $\beta$-equilibrium. To conclude, these are some approximations that we made, for simplicity, in order to sketch the model and obtain results, and may be easely included inside this formalism in a future work, just by adding the correspondent terms in the Lagrangian density. These three components are expressed together in the term proportional to $T^4$.

With the energy density and pressure expressed as above, we can use the equation
\begin{eqnarray}
\frac{d\left(\epsilon R^{3}\right)}{dR}=-3pR^{2} \label{eq:RRR}
\end{eqnarray}
and the definitions $\epsilon_{b}+\epsilon_{B}+\epsilon_{r}=\epsilon_{T}$ and $p_{b}+p_{B}+\frac{\epsilon_{r}}{3}=p_{T}$ to obtain
\begin{eqnarray}
\left(\frac{\partial\epsilon_{T}}{\partial\beta}\right)_{z,\phi}\frac{d\beta}{dt}+
\left(\frac{\partial\epsilon_{T}}{\partial z}\right)_{\beta,\phi}\frac{dz}{dt} \nonumber \\
+\left[\left(\frac{\partial\epsilon_{T}}{\partial\phi}\right)_{\beta,z} +
g_{\phi}\rho_{s}\right]\frac{d\phi}{dt}&=&-3\frac{\dot{R}}
{R}\left(p_{T}+\epsilon_{T}\right) , \;\;  
\label{ABCD}
\end{eqnarray}

\noindent
where $z=e^{\beta \mu_b}$ and $\beta=\frac{1}{T}$. Note that the index in derivatives are the variables that are left constant. 

Now we can complete the set of equations with the Friedman equation (the first equation written below) and get
\begin{eqnarray}
\dot{\mathcal{R}}^{2} & = & \frac{8\pi G}{3}\epsilon\mathcal{R}^{2},\label{eq:r^2}\\
\frac{d^{2}\phi}{dt^{2}}+3\frac{\dot{\mathcal{R}}}{\mathcal{R}}\frac{d\phi}{dt}+m_{\phi}^{2}\phi+\frac{\partial U}{\partial\phi} &=& g_{\phi}\rho_{s},\label{eq:eqphi}\\
\left(\frac{\partial\epsilon_{T}}{\partial\beta}\right)_{z,\phi}\frac{d\beta}{dt}+\left(\frac{\partial\epsilon_{T}}{\partial z}\right)_{\beta,\phi}\frac{dz}{dt} \nonumber \\
+\left[\left(\frac{\partial\epsilon_{T}}{\partial\phi}\right)_{\beta,z} +g_{\phi}\rho_{s}\right]\frac{d\phi}{dt}&=&-3\frac{\dot{\mathcal{R}}}{\mathcal{R}}\left(p_{T}+\epsilon_{T}\right),\;\;\label{eq:eqres3}\\
n_b & = & \frac{n_{b0}}{\mathcal{R}^{3}}  \  ,
\label{eq:nb_0232}
\end{eqnarray}
that are the equations for the hadronic phase of the Universe, the main objective of this work. 
The numerical solution of these equations will be published soon.
Here we have used the fact that the constant $k$, which should appear in the Friedman equation, can be taken as $0$, what leads to $\mathcal{R}=\frac{R}{R_0}$. This result will dispense us from the necessity of discovering the value of $R_0$, which demands the knowledge about the previous phases of the Universe. 

Although the complexity of the equations, we can get analytical results if we use some 
approximations in order to study the main characteristics of the solutions. 
First we consider only the term with $T^{4}$ in the energy density and pressure (as we are in the radiation-dominated 
Universe) 
and, through the equation (\ref{eq:RRR}), we get $T=\frac{T_0}{\mathcal{R}}$ and
\begin{equation}
\mathcal{R}=\sqrt{12,3\cdot10^{3}t+1}
\end{equation}
where the time is given in seconds. 
In a non-relativistic approximation
\begin{equation}
E_{i}\simeq m_{i}+\frac{1}{2}\frac{k^{2}}{m_{i}},  \  \  
\end{equation}
$\mu_{i}= m_{i}+\mu_{i}^{NR}$, and considering a very small contribution of the $\phi$ field (and of the vector fields), we
 get
\begin{eqnarray}
f_{i+}&\simeq&\frac{1}{e^{\beta\left(\frac{1}{2}\frac{k^{2}}{m_{i}}-\mu_{i}^{NR}\right)}+1},\quad i=p,n.\\f_{i-}&\simeq&\frac{1}{e^{\beta\left(\frac{1}{2}\frac{k^{2}}{m_{i}}+\mu_{i}^{NR}+2m_{i}\right)}+1}  \  \  .
\end{eqnarray}
Now, with the approximation
\begin{eqnarray}
\int_{0}^{\infty}\frac{x^{\lambda-1}dx}{z^{-1}e^{x}+1}\simeq \Gamma\left(\lambda\right)z,
\end{eqnarray}
where $\Gamma\left(\lambda\right)$ is the gamma function, we have as density of particles and anti-particles
\begin{eqnarray}
n_{i+}&=&\frac{2\left(2\pi m_{i}\right)^{\frac{3}{2}}}{\left(2\pi\right)^{3}}zT^{\frac{3}{2}}e^{-\frac{m_{i}}{T}},\\n_{i-}&=&\frac{2\left(2\pi m_{i}\right)^{\frac{3}{2}}}{\left(2\pi\right)^{3}}z^{-1}T^{\frac{3}{2}}e^{-\frac{m_{i}}{T}}.
\end{eqnarray}
So we can write
\begin{eqnarray}
n_i=\frac{2\left(2\pi m_{i}\right)^{\frac{3}{2}}}{\left(2\pi\right)^{3}}\left(z-z^{-1}\right)T^{\frac{3}{2}}e^{-\frac{m_{i}}{T}}. \label{eq:n_i}
\end{eqnarray}
With the condition $n_b=n_n+n_p=\frac{n_{b0}}{\mathcal{R}^3}=n_{b0}\left(\frac{T}{T_{0}}\right)^{3}$ and noting that $z-z^{-1}=2\sinh\left(\beta\mu_{b}\right)$, we get
\begin{eqnarray}
\mu_{b}=T\mathrm{arcsinh}\left(\alpha \right),
\end{eqnarray}
where
\begin{eqnarray}
\alpha=\frac{\left(2\pi\right)^{3}n_{b0}\left(\frac{T}{T_{0}}\right)^{3}e^{\frac{m_{n}}{T}}}{4\left(2\pi m_{n}T\right)^{\frac{3}{2}}\left(1+\left(\frac{m_{p}}{m_{n}}\right)^{\frac{3}{2}}e^{\frac{\left(m_{n}-m_{p}\right)}{T}}\right)}.  \  \ 
\end{eqnarray}
Then we have the densities given by
\begin{eqnarray}
n_{i+}&=&\frac{2\left(2\pi m_{i}\right)^{\frac{3}{2}}}{\left(2\pi\right)^{3}}T^{\frac{3}{2}}e^{-\frac{m_{i}}{T}+\mathrm{arcsinh}\left(\alpha\right)},\\n_{i-}&=&\frac{2\left(2\pi m_{i}\right)^{\frac{3}{2}}}{\left(2\pi\right)^{3}}T^{\frac{3}{2}}e^{-\frac{m_{i}}{T}-\mathrm{arcsinh}\left(\alpha\right)}.
\end{eqnarray}
The graphic of these expressions is shown in Fig. 1. As we can see, at high temperatures, 
the densities of the considered particles are very similar, but as
the temperature decreases, $\overline p$ and $\overline n$ decreases, and only protons and neutrons remain. The densities of $p$ and $n$, that at high temperatures are the same, present different values as the temperature becomes small. All these characterisitics 
appears in the results of \cite{raf2}, where the calculations are made with a different model. So, we may conclude, that even with
the approximations made, the final results are consistent.


\begin{figure}
\begin{centering}
\epsfxsize=8.cm
\epsffile{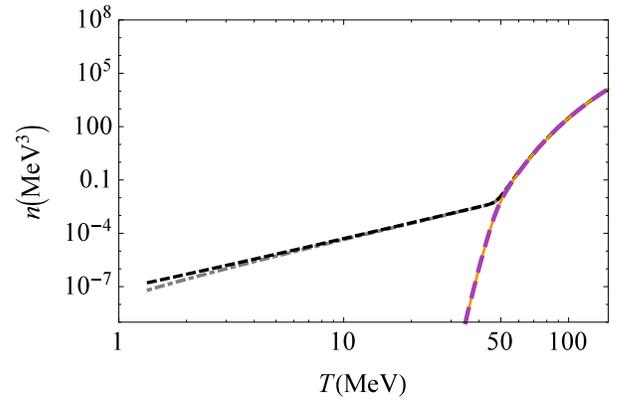}
\par\end{centering}

\protect\caption{Baryons and anti-baryons densities. The dashed (with smaller dash) line is the proton density, the dashed interspersed with points line is the neutron density, the continuous line is the anti-neutron density and the dashed (with biggest dash) line is the anti-proton density.}

\label{fg1}
\end{figure}

In this work we have studied the first second of the Universe (after $10^{-6}s$), that is, the hadronic phase. 
We obtained the analog of the Friedman set of equations, 
considering hadronic matter described by a relativistic mean-field model in the Robertson-Walker metric. This model opens 
possibilities for many variations and refinements, that can be made in a straightforward way.
In order to check the 
theory, a non-relativistic approximation has been made and consistent results have been obtained. The main characteristics
for the $p$, $n$, $\overline p$ and $\overline n$ have been described.

We would like to acknowledge S. B. Duarte and M. Malheiro for the very interesting discussions.


\end{document}